\pdfoutput=1

\documentclass[11pt,twoside,a4paper,cmspaper,final,collab]{cms-tdr}
\def\svnVersion{177391}\def\cmsCernNo{2012-152}\def\cmsCernDate{\today}\def\cmsMessage{Submitted to the Journal of High Energy Physics}

\begin{document}\cmsNoteHeader{QCD-10-021}

\hyphenation{had-ron-i-za-tion}
\hyphenation{cal-or-i-me-ter}
\hyphenation{de-vices}

\RCS$Revision: 73022 $
\RCS$HeadURL: svn+ssh://krabbert@svn.cern.ch/reps/tdr2/notes/QCD-10-021/trunk/QCD-10-021.tex $
\RCS$Id: QCD-10-021.tex.ex 73022 2011-08-02 12:24:41Z krabbert $
\graphicspath{{pdf/}{pdf/900GeV/}} % Note, the additional {} are mandatory!
\newlength{\mygraphicswidth}
\setlength{\mygraphicswidth}{0.50\textwidth}

\providecommand{\FASTJET}{{\textsc{fastjet}}\xspace}
\providecommand{\PROFESSOR}{{\textsc{professor}}\xspace}
\providecommand{\etaphi}{\ensuremath{\eta}-\ensuremath{\phi}\xspace}
\providecommand{\rhop}{\ensuremath{\rho'}\xspace}
\providecommand{\rhopm}{\ensuremath{\left<\rho'\right>}\xspace}
\providecommand{\hftwo}{\hspace*{\fill}}
\providecommand{\rbthm}{\rule[-2ex]{0ex}{5ex}}
\providecommand{\rbthr}{\rule[-1.7ex]{0ex}{5ex}}
\providecommand{\rbtrm}{\rule[-2ex]{0ex}{5ex}}
\providecommand{\rbtrr}{\rule[-0.8ex]{0ex}{3.2ex}}

\cmsNoteHeader{QCD-10-021}% This is over-written in the CMS environment: useful as preprint no. for export versions
\title{Measurement of the underlying event activity in pp collisions
  at $\sqrt{s} = 0.9$ and $7\TeV$ with the novel jet-area/median
  approach}% Force line breaks with \\

\date{\today}

\abstract{
  The first measurement of the charged component of the underlying
  event using the novel ``jet-area/median'' approach is presented for
  proton-proton collisions at centre-of-mass energies of 0.9 and
  7\TeV. The data were recorded in 2010 with the CMS experiment
  at the LHC\@. A new observable, sensitive to soft particle
  production, is introduced and investigated inclusively and as a
  function of the event scale defined by the transverse momentum of
  the leading jet. Various phenomenological models are compared to
  data, with and without corrections for detector effects.  None of
  the examined models describe the data satisfactorily.
}

\hypersetup{%
  pdfauthor={CMS Collaboration},%
  pdftitle={Measurement of the underlying event activity in pp collisions
     at sqrt(s) = 0.9 and 7 TeV with the novel jet-area/median approach},%
  pdfsubject={CMS},%
  pdfkeywords={CMS, physics, QCD, underlying event, jet area}
}

\maketitle %maketitle comes after all the front information has been supplied

\clearpage
\section{Introduction}
\label{sec:introduction}

In the theoretical description of nondiffractive inelastic
proton-proton collisions, the main momentum transfer occurs between
only two partons. This simple picture is refined by the inclusion of
radiative effects in the form of initial- and final-state radiation.
In addition, the primary interaction is accompanied by the production
of further particles in multiple-parton interactions (MPIs) and in the
hadronisation of the beam-beam remnants. The extra activity in a
collision, which cannot be uniquely separated from initial- and
final-state radiation, is referred to as the underlying event (UE)\@.

Monte Carlo event generators simulate the UE based on phenomenological
models, which have been tuned to the data of various collider
experiments, taking into account the dependence of the UE on the
centre-of-mass energy. The observation of substantial deviations of
the predictions from the data, in particular when extrapolating to
different centre-of-mass energies, emphasises the need for
measurements of the UE at different
energies~\cite{Affolder:2001xt,Acosta:2004wqa,Aaltonen:2010rm,
  CMS-PAPERS-QCD-10-010,Aad:2010fh,Aad:2011qe}.  Retuned models allow
for more precise measurements of observables based on jets or relying
on isolation cones, for example in diphoton events from QCD processes
or decays of the Higgs boson.

An unambiguous association of a specific particle to the reaction from
which it originates is impossible. The investigation of the UE
therefore requires a physically motivated separation of hard and soft
contributions through the definition of phase-space regions that are
dominated by either the hard or soft component of a collision.
Traditionally, this is done by geometrically subdividing an event into
different regions (``towards'', ``away'', and ``transverse'') with
respect to the jet or particle leading in transverse momentum \pt. At
the same time the \pt of the leading object is defined to be the
so-called ``event scale'', i.e.\ a measure of the momentum transfer in
the hard partonic scattering. Studies using this approach were
performed at the
Tevatron~\cite{Affolder:2001xt,Acosta:2004wqa,Aaltonen:2010rm} and the
Large Hadron Collider
(LHC)~\cite{CMS-PAPERS-QCD-10-010,Aad:2010fh,Aad:2011qe}.

A new technique based on the transverse momentum density per jet area,
the jet-area/median approach, was proposed in~\cite{Cacciari:2009dp}.
The jet area covered by a jet, the ``catchment
area''~\cite{Cacciari:2008gn}, is determined in the plane of
pseudorapidity $\eta$ versus azimuthal angle $\phi$ as defined in
Section~\ref{sec:observable}. The exact size and shape of the area
must be sensitive to the event-by-event fluctuating soft hadronic
activity of the UE\@. The most widely used jet algorithm at the LHC,
the anti-\kt jet algorithm~\cite{Cacciari:2008gp}, is unsuited for
such an analysis and is replaced by the \kt
algorithm~\cite{Catani:1991hj,Catani:1992zp,Ellis:1993tq,Cacciari:2005hq}.
The separation of the soft from the hard component of a collision is
performed event by event by using the median of the distribution of
transverse momentum densities of all jets in an event.

The data analysed in this study were collected with the Compact Muon
Solenoid (CMS) detector at centre-of-mass energies of $0.9$ and
$7$\TeV during the early LHC running in 2010, in which the
contamination of events by additional proton-proton collisions in or
close to the same bunch crossing, so-called pileup collisions, is very
small. This jet area technique can be exploited to correct jet
energies for pileup contamination in other measurements. The present
paper is the first publication applying this new method in a collider
experiment.

In the following, Section~\ref{sec:observable} defines the
UE-sensitive observable based on jet areas.
Section~\ref{sec:detector_data_event} describes the experimental setup
for data collection, triggering, vertex reconstruction, and event
selection.  Section~\ref{sec:mc} gives details on the phenomenological
models used for event generation and on the detector simulation. The
event reconstruction and the track and jet selections are explained in
Section~\ref{sec:reconstruction} and are followed by a description of
the unfolding technique in Section~\ref{sec:unfolding}.
Sections~\ref{sec:systematics} and~\ref{sec:results} present the
derivation of the measurement uncertainties and the final results,
which are then summarised in Section~\ref{sec:summary}.

\section{Definition of the observable}
\label{sec:observable}

CMS uses a right-handed coordinate system, with the origin at the
nominal interaction point, the $x$ axis pointing to the centre of the
LHC, the $y$ axis pointing up (perpendicular to the LHC plane), and
the $z$ axis along the anticlockwise-beam direction. The polar angle
$\theta$ is measured from the positive $z$ axis and the azimuthal
angle $\phi$ is measured in the $x$-$y$ plane. The pseudorapidity
$\eta$ is then defined as $\eta = -\ln\tan(\theta/2)$.

The adopted standard for jet clustering in the LHC experiments is the
anti-\kt jet algorithm~\cite{Cacciari:2008gp}. Although it follows a
sequential recombination procedure, the jets leading in \pt resemble
in shape the jets reconstructed using algorithms with fixed cone
sizes~\cite{Sterman:1977wj} because it starts clustering with the
hardest (highest \pt) objects. Hence, it is less sensitive to details
of the distribution of softer objects in an event and less suited for
an investigation of the UE\@. In contrast, the \kt jet algorithm
clusters the softest objects first, trying to undo the effects of
parton showering~\cite{Catani:1991hj,Catani:1992zp,Ellis:1993tq}. This
approach to jet clustering leads to nonuniform catchment areas of \kt
jets, which can be evaluated by applying the active area clustering
technique as described in~\cite{Cacciari:2008gn}. In this analysis of
the UE, jets are reconstructed using the \kt algorithm with a distance
parameter $R$ of $0.6$ as implemented in the \FASTJET
package~\cite{Cacciari:2005hq,Cacciari:2011ma}, which at the same time
performs the jet-area determination.  For this purpose, the event in
question is overlaid with a uniform grid of artificial, extremely soft
``ghost particles'' in the $\eta$-$\phi$ plane as indicators of a
jet's domain of influence or catchment area. They are fed into the jet
algorithm together with the measured tracks or charged particles but
without impact on the reconstructed physical jets. This is guaranteed
by the use of an infrared- and collinear-safe jet algorithm and the
smallness of the transverse momentum of the ghost particles, which is
of the order of $10^{-100}\GeV$. The number of ghosts, $N_j^\text{
  ghosts}$, clustered into a jet $j$ is then a measure of the jet area
$A_j$:

\begin{equation}
  A_j = \frac{N_j^\text{ghosts}}{\rho^\text{ghosts}} =
  \frac{N_j^\text{ghosts}}{N_\text{tot}^\text{ghosts}}{A_\text{tot}}\ ,
\end{equation}

where $\rho^\text{ghosts}$ is the ghost density, defined as the total
number of ghosts $N_\text{tot}^\text{ghosts}$ divided by the total area
$A_\text{tot}$ within the acceptance. In this study, $A_\text{tot}$ is
set equal to $8\pi$ according to the ranges of $0 \leq \phi < 2\pi$
and $|\eta| \leq 2$. In order to limit boundary effects, the
directions of the jets axes are restricted to $|\eta| < 1.8$ while
tracks are used up to $|\eta| = 2.3$. The distribution of ghosts
extends up to $|\eta| = 5$. Here it is important to note that empty
areas within the acceptance are covered by jets which consist solely
of ghost particles. These ``ghost jets'' have a well-defined area but
vanishing transverse momentum.

A measure of the soft activity in an event is then given by the median
$\rho$ of the distribution of the jet transverse momentum per jet area
for all jets in an event~\cite{Cacciari:2009dp}:

\begin{equation}
  \rho = \underset{j\,\in\,\text{jets}}{\text{median}}
  \left\{{\frac{p_{\mathrm{T}j}}{A_j}}\right\}\,.
\end{equation}

The choice of the median is motivated by its robustness to outliers in
the distribution. These outliers are in particular the leading jets
originated by the hard partonic interaction. The observable $\rho$
thus naturally isolates UE contributions by assuming that the majority
of the event is either empty or dominated by soft contributions and
that the hard component of the interaction is well contained within
the leading jets. In contrast to the conventional approach, no
explicit geometrical subdivision of the event is necessary. The
separation of the hard and soft components is done event by event
through the area clustering for the \kt algorithm and the use of the
median.  An advantage of this novel method is that it can easily be
extended to event topologies other than the minimum-bias events
investigated here.

In the proposal for a measurement of $\rho$ in collider
experiments~\cite{Cacciari:2009dp}, no kinematic selection was imposed
on the input particles for the jet clustering. Unfortunately, this is
not realistic experimentally because of threshold effects and a limited
detector acceptance. Typically in minimum-bias events with a low
average charged-particle multiplicity, large parts of the detector do
not contain any physical particles and are therefore covered by ghost
jets. As each ghost jet contributes one entry at $p_{\mathrm{T}j}/A_j
= 0$, events with a majority of ghost jets have $\rho = 0$,
corresponding to zero UE activity. In order to increase the
sensitivity to low UE activity, an adjusted observable \rhop is
introduced, which takes into account only jets containing at least one
physical particle:

\begin{equation}
  \rhop = \underset{j\,\in\,\text{physical jets}}
  {\text{median}}\left\{{\frac{p_{\mathrm{T}j}}{A_j}}\right\}\cdot C\,.
\end{equation}

Here, the event occupancy $C$, defined as the area $\sum_{j}{A_{j}}$
covered by physical jets divided by the total area $A_\text{tot}$, is a
measure of the ``nonemptiness'' of an event. While in $\rho$ the ghost
jets account for empty regions in the detector in the derivation of
the median, the scaling factor $C$ has a similar effect on \rhop by
shifting events with low activity towards smaller values of \rhop.  In
the limit of full coverage of the detector with physical jets, $\rho$
and \rhop are identical.

\section{Detector description and event selection}
\label{sec:detector_data_event}

The central feature of the CMS apparatus is a superconducting solenoid
of 6\unit{m} internal diameter. Within the field volume are a
silicon pixel and strip tracker, a crystal electromagnetic
calorimeter, and a brass/scintillator hadron calorimeter. Muons are
measured in gas-ionisation detectors embedded in the steel return
yoke. Extensive forward calorimetry complements the coverage provided
by the barrel and endcap detectors. In the following, only the parts
of the detector that are most important for this analysis will be
presented.  A more detailed description can be found in
Ref.~\cite{CMS-DET}.

The inner tracker measures charged particles within the pseudorapidity
range $|\eta| < 2.5$. It consists of 1440 silicon pixel and 15\,148
silicon strip detector modules and is located in the $3.8\unit{T}$
field of the superconducting solenoid. It provides an impact parameter
resolution of ${\sim} 15\mum$ and a transverse momentum
resolution of about $1.5\%$ for $100\GeV$ particles.

Two subsystems of the first-level trigger system are used in this
analysis: the Beam Pick-up Timing for eXperiments (BPTX) and the Beam
Scintillator Counters (BSC). The two BPTX devices, which are installed
around the beam pipe at a distance of $\pm 175\unit{m}$ from the
interaction point, are designed to provide precise information on the
structure and timing of the LHC beams with a time resolution better
than $0.2\unit{ns}$.  The two BSCs, each consisting of a set of 16
scintillator tiles, are located along the beam line on each side of
the interaction point at a distance of $10.86\unit{m}$ and are
sensitive in the range $3.23 < |\eta| < 4.65$. They provide
information on hits and coincidence signals with an average detection
efficiency of $96.3\%$ for minimum-ionising particles and a time
resolution of $3\unit{ns}$.

For an analysis of the UE, only data with not more than one collision
per bunch crossing, i.e.\ without pileup, are suitable. Therefore data
taken during periods with low instantaneous luminosity, between March
and August $2010$, at centre-of-mass energies of $0.9$ and $7\TeV$ are
used.

The high-level trigger selection requires at least one track to be
reconstructed in the silicon pixel detector with a minimum transverse
momentum of $0.2\GeV$. This high-level trigger path is seeded by the
BPTX and BSC level-1 triggers. In order to minimise the contamination
caused by additional interactions within the same LHC bunch crossing,
only events with exactly one reconstructed vertex are used in this
analysis. The position of this vertex must be fitted from at least
four tracks and its $z$ component must lie within $10\unit{cm}$ of the
centre of the reconstructed luminous region for the given data-taking
run~\cite{CMS-PAPERS-TRK-10-001}.
The effect of pileup collisions that remained
undetected because of inefficiencies in the primary vertex
reconstruction is estimated to be negligible compared to other sources
of systematic uncertainty.

Even though this analysis contains data taken at two different
centre-of-mass energies, all event selection and trigger criteria are
identical throughout to guarantee compatibility of the results and
consistency with the conventional UE
measurement~\cite{CMS-PAPERS-QCD-10-010}.

\section{Event generators and simulation}
\label{sec:mc}

The generator predictions that are compared with the data were
produced with three different tunes of the \PYTHIA program
version~6.4.22~\cite{Sjostrand:2006za}, and one from \PYTHIA
version~8.145~\cite{Sjostrand:2007gs}. The matrix elements chosen for
the event generation reflect the minimum-bias event selection in data.
A simulation of the CMS detector, based on the \GEANTfour
package~\cite{Agostinelli:2002hh}, is applied. As Monte Carlo methods
are used in both steps, we refer to ``generator'' when particle-level
generator information is concerned, while ``simulation'' refers to a
simulation of the CMS detector response.

The \PYTHIA 6 tune D6T~\cite{Field:2008zz,Field:2010su} was the
default tune within the CMS Collaboration prior to the LHC operation.
It is based on the CTEQ6L1~\cite{Pumplin:2002vw} parton distribution
functions (PDFs) and was tuned to describe measurements of the UA5
Collaboration at the S\Pp\Pap{}S collider and the Tevatron
experiments.

As a consequence of the higher particle multiplicities observed in LHC
collision data at $0.9\TeV$ and
$7\TeV$~\cite{Aamodt:2010ft,Aamodt:2010pp,Aad:2010rd,CMS-PAPERS-QCD-09-010,CMS-PAPERS-QCD-10-006}
compared to existing model predictions, the new tunes
Z1~\cite{Field:2010bc} and Z2 were developed. Both tunes employ a new
model for MPIs and a fragmentation function derived with the
\PROFESSOR~\cite{Buckley:2009bj} tool, as well as \pt-ordered parton
showers. The main difference between the two tunes is the usage of the
CTEQ5L PDFs~\cite{Lai:1999wy} in Z1 and the CTEQ6L1
PDFs~\cite{Pumplin:2002vw} in Z2. Using different PDF sets requires
the adjustment of the parameter that defines the minimal momentum
transfer in MPIs in order to keep the cross section of the additional
scatterings constant. Tune 4C~\cite{Corke:2010yf} of \PYTHIA version 8
also uses a new MPI model, which is interleaved with parton showering,
and the CTEQ6L1 PDFs.

During simulation and reconstruction, the simulated samples take into
account an imperfect alignment as well as nonoperational channels of
the tracking system.

\section{Reconstruction}
\label{sec:reconstruction}

For the purpose of measuring the soft activity of the UE, the data are
analysed down to very small transverse momenta of $0.3\GeV$,
exploiting the capabilities of the CMS tracking detectors. A potential
neutral component of the UE, measurable only with the calorimeters, is
neglected. Consequently, the jet clustering is performed on
reconstructed tracks either from data or simulated events (track
jets), and also on stable charged particles in generator events
(charged-particle jets). Generator particles with mean lifetimes
$\tau$ such that $c\tau > 10\mm$ are considered to be stable.

\subsection{Track selection}

The performance and technical details of the CMS tracking with first
collision data is described in~\cite{CMS-PAPERS-TRK-10-001}. The track
selection of this analysis follows that of the conventional UE
measurement as discussed in~\cite{CMS-PAPERS-QCD-10-010}. In detail,
the following criteria are applied:

\begin{itemize}
\item{high-purity track quality~\cite{CMS-PAPERS-TRK-10-001};}
\item{transverse momentum $\pt > 0.3\GeV$;}
\item{pseudorapidity $|\eta| < 2.3$;}
\item{transverse impact parameter $d_{xy} < 0.2\cm$;}
\item{longitudinal impact parameter $d_z < 1\cm$;}
\item{relative track \pt uncertainty $(\sigma_{\pt}/\pt) \cdot \max
    (1,\chi^{2}/N_\mathrm{dof}) <0.2$, where $N_\mathrm{dof}$ denotes
    the number of degrees of freedom in the track fit.}
\end{itemize}

These impact parameters are determined with respect to the primary
vertex.

\subsection{Charged generator particles}

The influence of the detector on a particular observable is estimated
by comparing the predictions, as given by a particle generator, before
and after detector simulation, including trigger effects. To achieve a
good correspondence to the track selection, the generated stable
charged particles are required to satisfy $\pt > 0.3\GeV$ and $|\eta|
< 2.3$. This minimum transverse momentum threshold and the restriction
to charged particles significantly reduces the number of particles
entering the clustering process.

\subsection{Jet selection}

No further selection on the transverse momenta of the jets is imposed.
Because of the selection criteria on the input objects, however, they
are implicitly restricted to be larger than $0.3\GeV$. To avoid
boundary effects in the jet-area determination, the absolute
pseudorapidity of the jet axis is required to be smaller than $1.8$,
which is to be compared with $|\eta| < 2.3$ for the input objects.

\section{Detector unfolding}
\label{sec:unfolding}

In order to compare data with theoretical predictions, the measurement
must be corrected for detector response and resolution effects. In
abstract terms, the connection between a true and the reconstructed
distribution is expressed by a folding integral, which must be
inverted to correct for the detector effects. Commonly, this procedure
is referred to as unfolding or deconvolution.  The technique adopted
here to unfold the \rhop distribution is the iterative Bayesian
approach~\cite{D'Agostini:1994zf} as implemented in the RooUnfold
framework~\cite{Adye:2011ru}. For this method the relevant
distributions of a given observable are analysed before and after
detector simulation and the detector response is expressed as a
response matrix.  To improve the statistical stability of the
unfolding procedure, a wider binning and a reduced \rhop range are
used compared to the uncorrected distributions.

It is found that the response matrices derived from different event
generator tunes yield different results after the unfolding of the
data distribution, which is a consequence of the difference in track
multiplicities of the tunes. The tune Z2, which yields the best
description of track-based observables, is used to unfold detector
effects, while the others are employed only to derive the systematic
uncertainties arising from this procedure.

\section{Systematic uncertainties}
\label{sec:systematics}

The following sources of systematic uncertainties are considered: %
the trigger efficiency bias, %
the influence of the track selection, %
track misreconstruction and the reconstruction efficiency, %
the track jet \pt response, %
nonoperational tracker channels, %
and the tune dependence in the unfolding procedure.

Since most of the effects are found to be \rhop-dependent, suitable
parametrisations are chosen to quantify them. From these
parametrisations, the uncertainties are derived bin by bin by adding
the different effects in quadrature. For variations in the
requirements, for example from decreasing and increasing the track \pt
requirement, symmetrised uncertainties are derived in the form of the
average of the observed absolute deviations from the baseline
scenario. Representative numbers for the uncertainties are summarised
in Table~\ref{tab:sys_table} apart from the trigger efficiency bias,
which is found to be negligible, since the event selection criterion
of at least four tracks required for a well reconstructed primary
vertex is stricter than the trigger condition.

The only track selection criterion identified to have a significant
impact on the observable is the minimal track \pt. Varying the
threshold value of $300\MeV$ by ${\pm}10\%$ induces a systematic
uncertainty on the \rhop distribution of about 2.0\% at 7\TeV and
3.0\% at 0.9\TeV. For the lowest \rhop bins, the effect increases
dramatically to ${\pm}15\%$ at 7\TeV and ${\pm}16\%$ at 0.9\TeV.

The potential mismatch between the number of reconstructed tracks and
the number of charged particles is estimated from simulated events to
be $5\%$. A similar number is found for the reconstruction efficiency
of nonisolated muons in data~\cite{CMS-PAS-TRK-10-002}. To quantify
the influence of the tracking efficiency on \rhop, a random track from
an independent sample is added to the analysed sample with a
probability of $5\%$ per existing track. Thus, the kinematic variables
of the additional track follow the distributions predicted by the
simulation.  The effect of dropping each track with a probability of
$5\%$ is also studied. The total resulting influence on \rhop for
adding false or losing real tracks is found to be around 0.5\% in most
bins.

The response of the track jet \pt measurement compared to
charged-particle jets is another source of uncertainty. It is studied
by shifting the \pt of each jet in the events by ${\pm}1.7\%$. This
number corresponds to the width of the transverse momentum response
distribution when comparing jets from generated charged particles and
their corresponding reconstructed track jets. As expected in the case
of systematically increased transverse momenta, the \rhop spectrum is
shifted towards higher values and vice versa. The magnitude of the
effect is dependent on \rhop and ranges from about $4\%$ for small
\rhop to about $2.0\%$ for large \rhop at $\sqrt{s} = 7\TeV$. In the
case of $\sqrt{s} = 0.9\TeV$, the effect is more pronounced but it
remains smaller than $5\%$.

Further sources of uncertainties are nonoperational tracker channels
and imperfect alignment of the tracker components. These effects are
studied by means of a special simulated data set, which assumes
perfect alignment and all channels functional. Small values of \rhop
are affected most, with a total systematic uncertainty of 1.0\% at
7\TeV and 2.5\% at 0.9\TeV on average.

The uncertainty arising from the response matrix in the unfolding
procedure is evaluated by investigating the differences in the
response in the different tunes. The measured distribution is unfolded
with all four response matrices, and the average deviation of the D6T,
Z1, and 4C results from those obtained with the Z2 tune are taken as
the systematic uncertainty which, amounts to roughly 4\% at 7\TeV and
9\% at 0.9\TeV, increasing for higher \rhop values.

\begin{table}[ht]
  \topcaption{Summary of systematic uncertainties on the \rhop
    distributions (in percent).}
  \centering
  \begin{tabular}{l|cc|cc}
    \hline
    \multirow{2}{*}{Systematic effect}
    & \multicolumn{2}{c|}{$\sqrt{s} = 0.9\TeV$} &
    \multicolumn{2}{c}{$\sqrt{s} = 7\TeV$} \rbtrr\\
    & typ.\ size & max.\ size & typ.\ size & max.\ size \rbtrr\\
    \hline
    Track selection
    & $\pm 3.0$ & $\pm  16$ (low \rhop) & $\pm 2.0$ & $\pm  15$ (low \rhop) \rbtrr\\
    Track reconstruction
    & $\pm 0.5$ & $\pm 3.0$ (low \rhop) & $\pm 0.5$ & $\pm 2.5$ (low \rhop) \rbtrr\\
    Track-jet \pt response
    & $\pm 4.0$ & $\pm 5.0$ (low \rhop) & $\pm 2.0$ & $\pm 4.0$ (low \rhop) \rbtrr\\
    Nonoperational tracker channels
    & $\pm 2.5$ & $   -3.0$ (low \rhop) & $\pm 1.0$ & $  +1.5$ (low \rhop) \rbtrr\\
    Unfolding \& tune dependence
    & $\pm 9$   & $\pm 10$  (high \rhop)& $\pm 4$   & $\pm 16$  (high \rhop)\rbtrr\\
    \hline
  \end{tabular}
  \label{tab:sys_table}
\end{table}

\section{Results}
\label{sec:results}

As in conventional UE measurements it is possible for the \rhop
observable to be investigated not only inclusively but also as a
function of the hardness of an event, which is given by the ``event
scale''. In the conventional approach, this scale is usually defined
by the transverse momentum of either the hardest track or hardest jet.
In the present study, the natural choice for the event scale is the
transverse momentum of the jet leading in \pt within the
acceptance. In the next two subsections the inclusive and the
event-scale-dependent results on \rhop are presented without
correction for detector effects. The unfolded results follow in
Subsection~\ref{sec:unfoldedresults}.

\subsection{Inclusive measurement}

\begin{figure}
  \includegraphics[width=\mygraphicswidth]{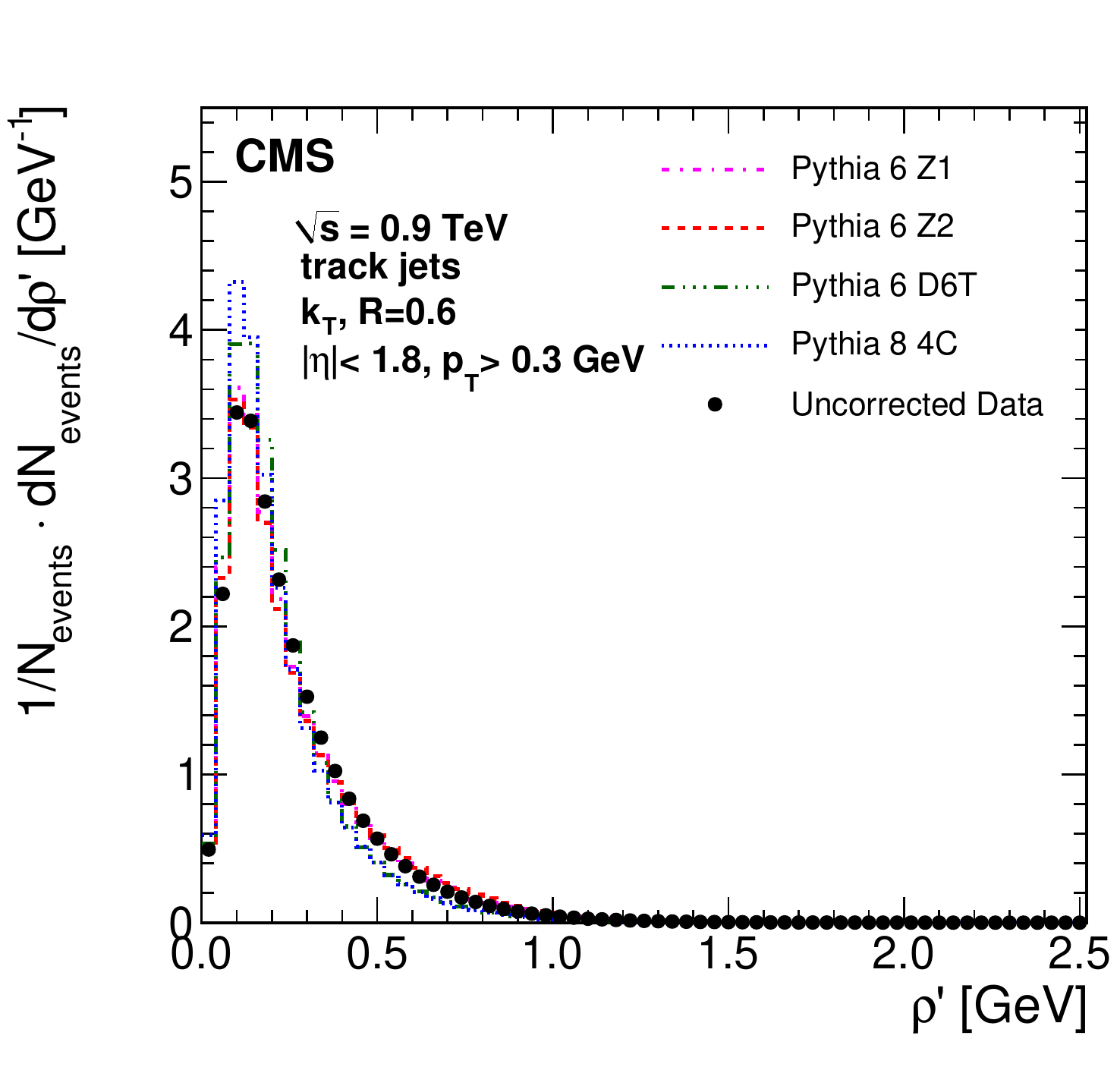}\hftwo%
  \includegraphics[width=\mygraphicswidth]{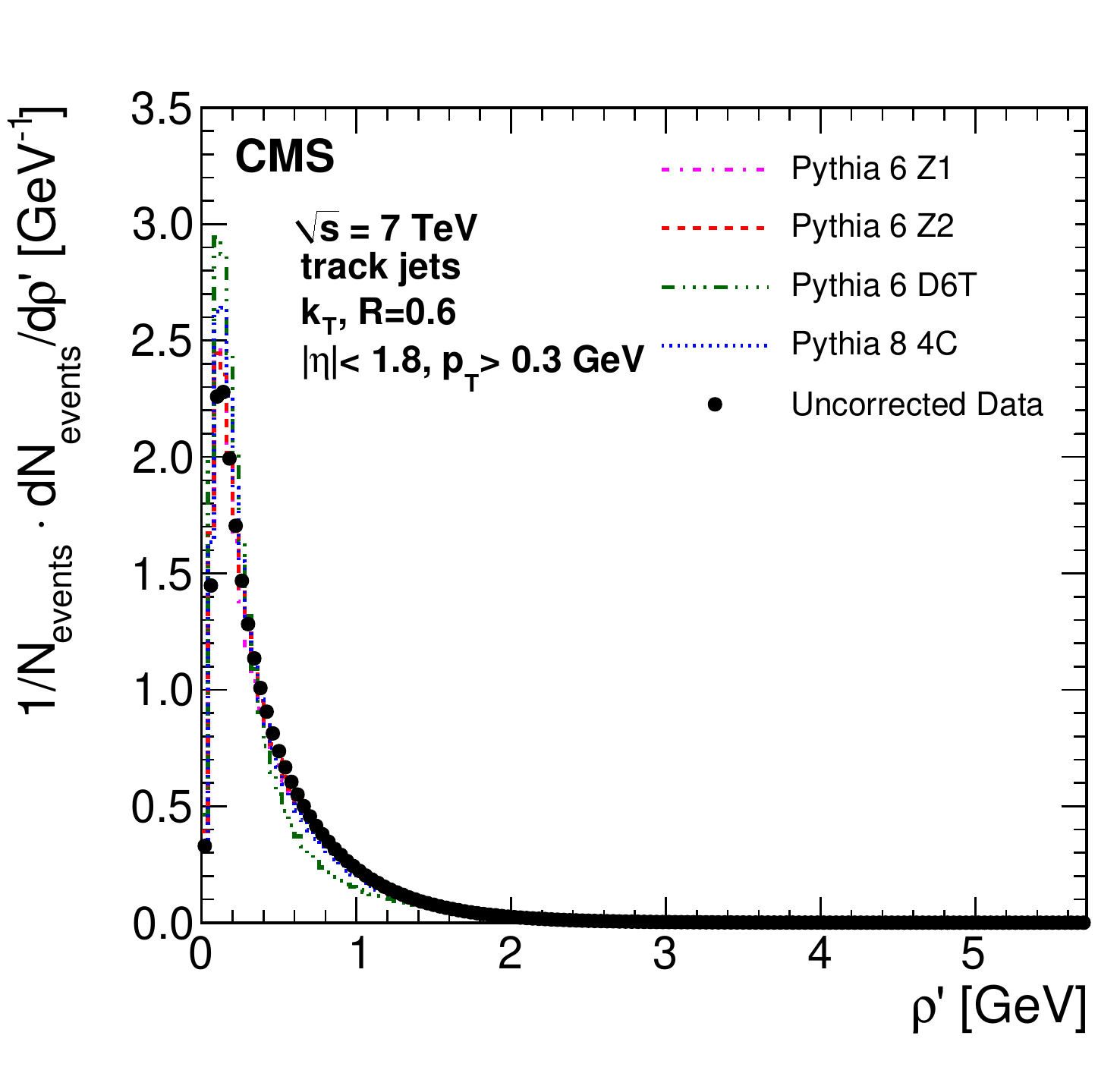}
  \includegraphics[width=\mygraphicswidth]{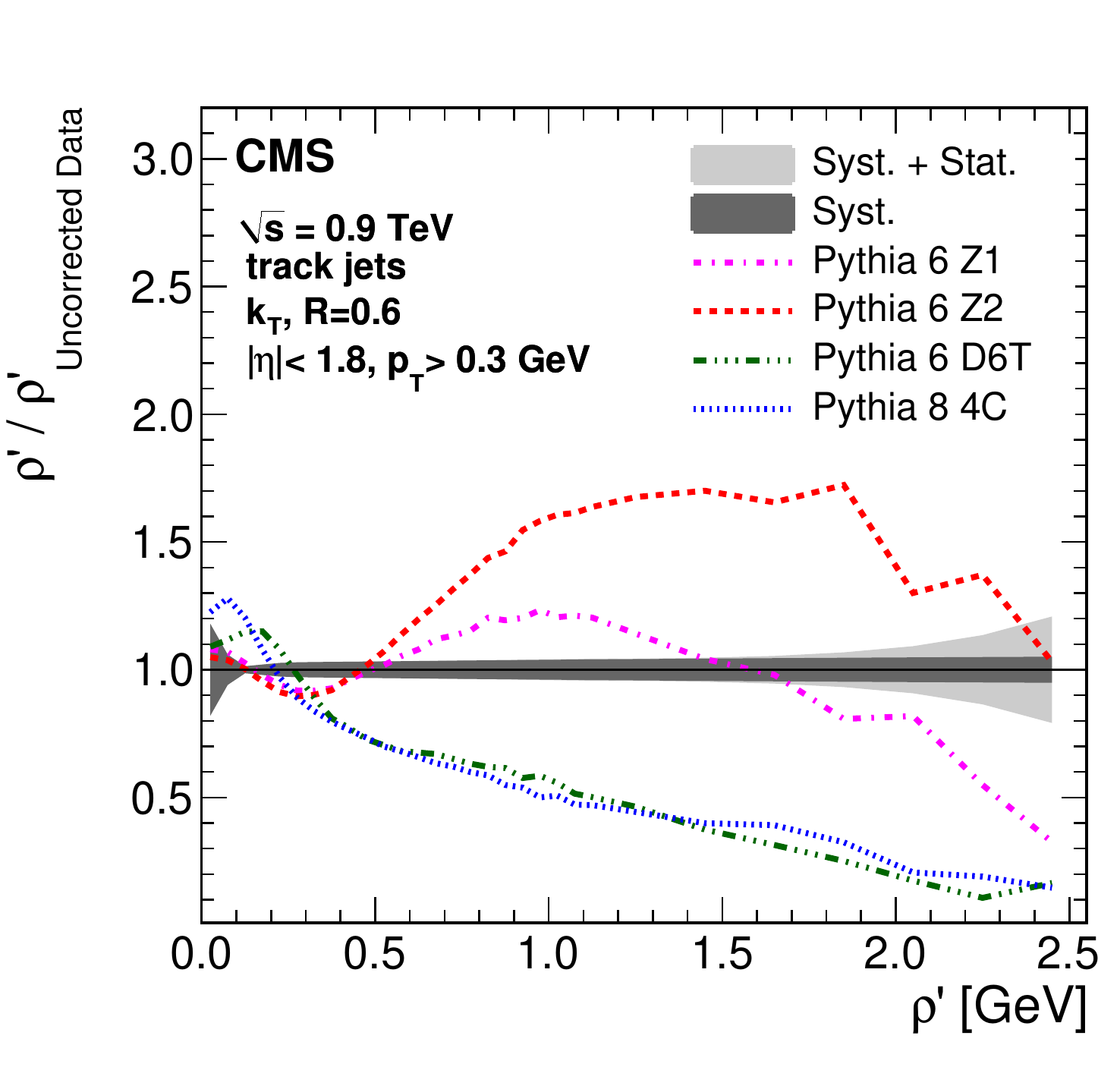}\hftwo%
  \includegraphics[width=\mygraphicswidth]{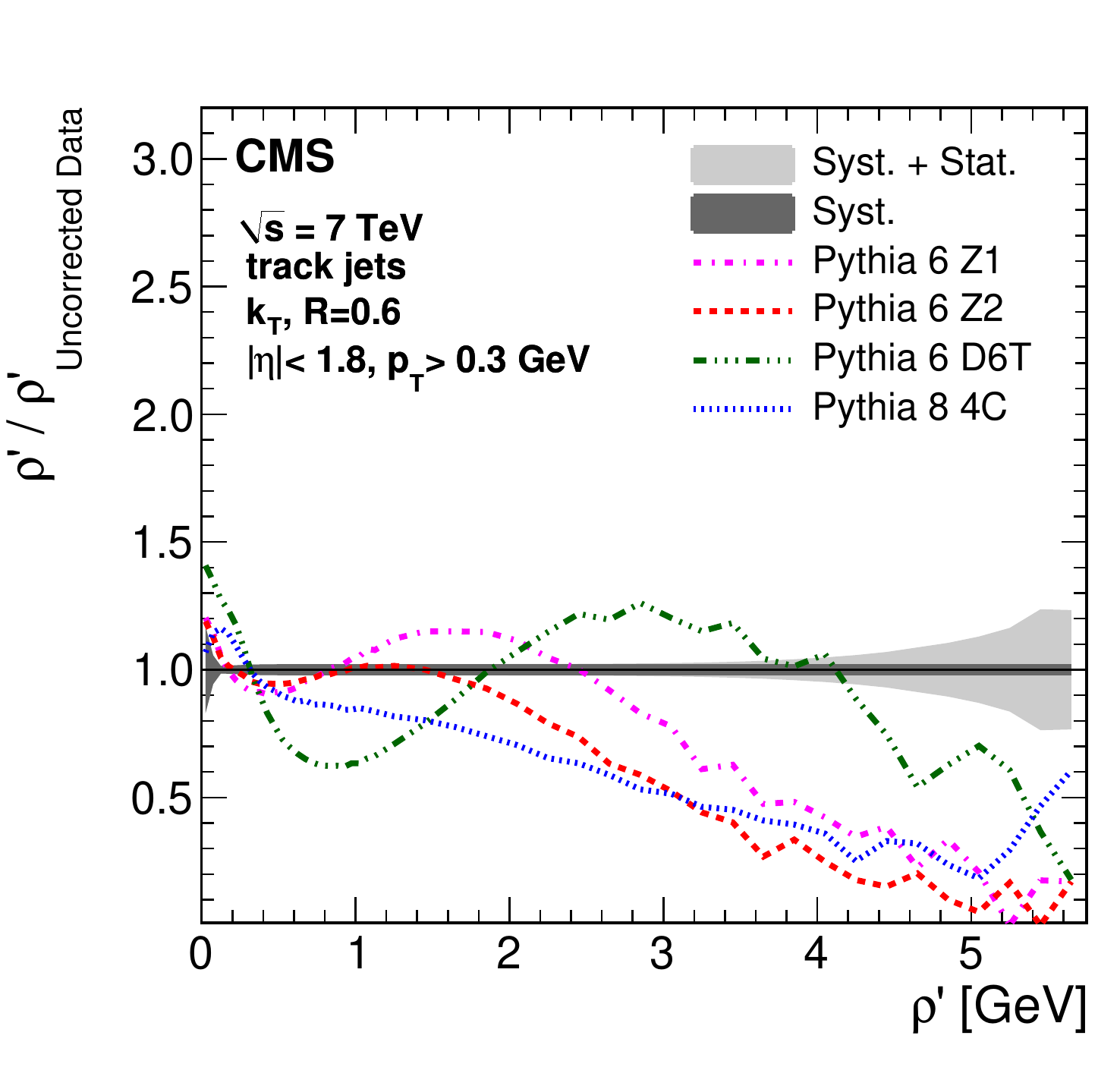}
  \caption{Uncorrected inclusive \rhop distributions for data and
    simulation (upper row), and ratios of the \PYTHIA~6 tunes Z1, Z2,
    D6T, and the \PYTHIA~8 tune 4C relative to data (lower row) at
    $\sqrt{s} = 0.9\TeV$ (left) and $\sqrt{s} = 7\TeV$ (right). The
    dark grey shaded band corresponds to the systematic uncertainty
    and the light grey shaded band to the quadratic sum of the
    systematic and statistical uncertainty. The reach
    in \rhop is different at the two centre-of-mass energies.}
  \label{Fig:kt6TrackJetsWithArearho}
\end{figure}

Figure~\ref{Fig:kt6TrackJetsWithArearho} shows the uncorrected
inclusive \rhop distributions for data in comparison to the \PYTHIA~6
tunes Z1, Z2, D6T, and the \PYTHIA~8 tune 4C\@. The distributions are
normalised to the observed number of events. All predictions deviate
significantly from the measurements at both centre-of-mass energies,
in particular for \rhop values larger than about 0.5\GeV. At $\sqrt{s}
= 0.9\TeV$ \PYTHIA~6 Z2 overshoots the data while \PYTHIA~6 D6T and
\PYTHIA~8 4C are systematically too low. In comparison, \PYTHIA~6 Z1
is closer to the measurement with some overestimation in the range of
\rhop from 0.5 to 1.5\GeV at $\sqrt{s} = 0.9\TeV$ and a similar
behaviour from 1.0 to 2.0\GeV at $\sqrt{s} = 7\TeV$. For higher \rhop,
Z1 undershoots the data. While \PYTHIA~8 4C continues to exhibit too
little UE activity at the higher centre-of-mass energy of 7\TeV,
\PYTHIA~6 D6T describes the data somewhat better.  \PYTHIA~6 Z2
changes from severely overestimating the UE to an underestimation at
7\TeV, hinting at a problem with the energy dependence of the UE in
this tune.

\subsection{Event scale dependence}

\begin{figure}
  \includegraphics[width=\mygraphicswidth]{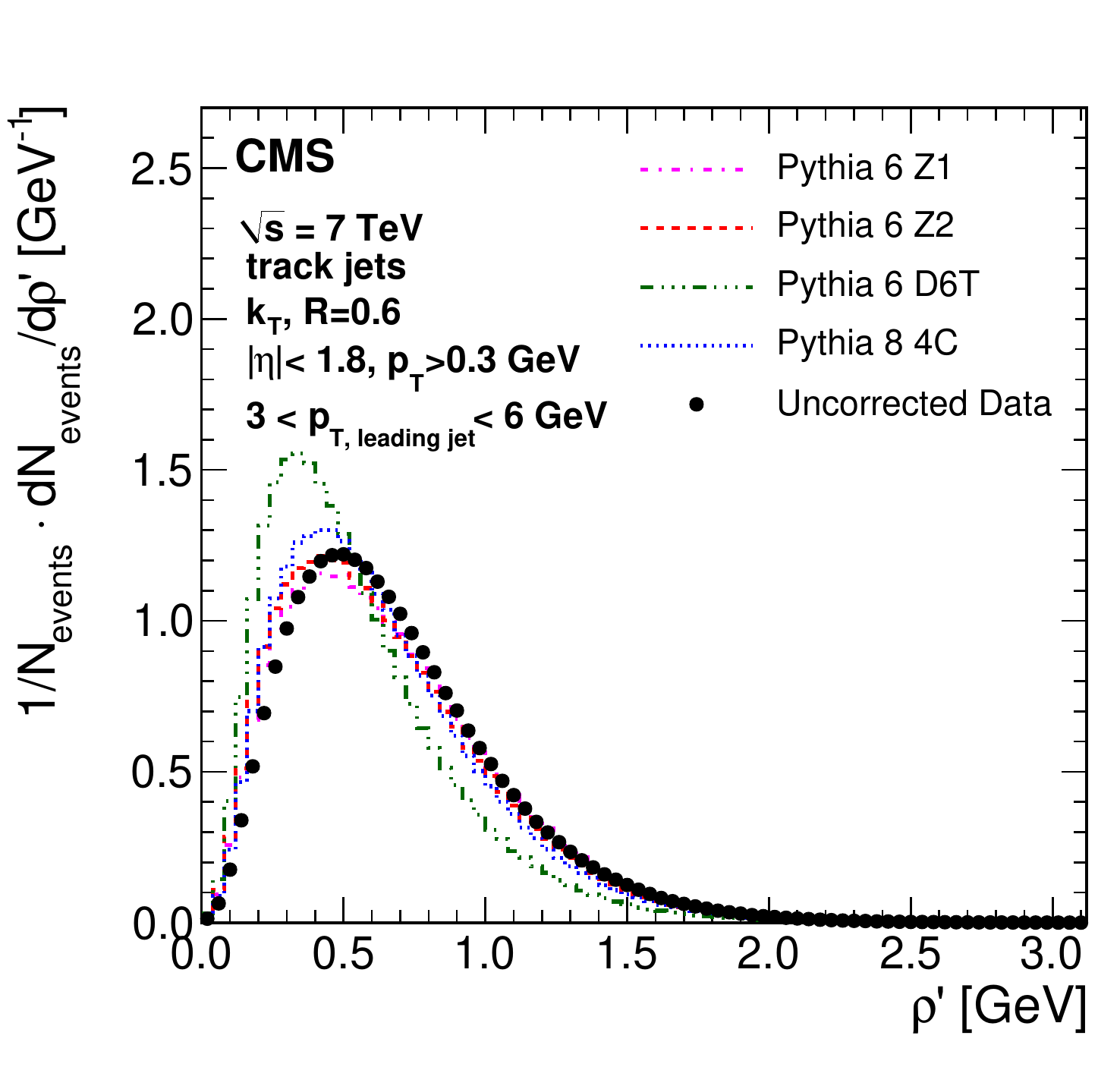}\hftwo%
  \includegraphics[width=\mygraphicswidth]{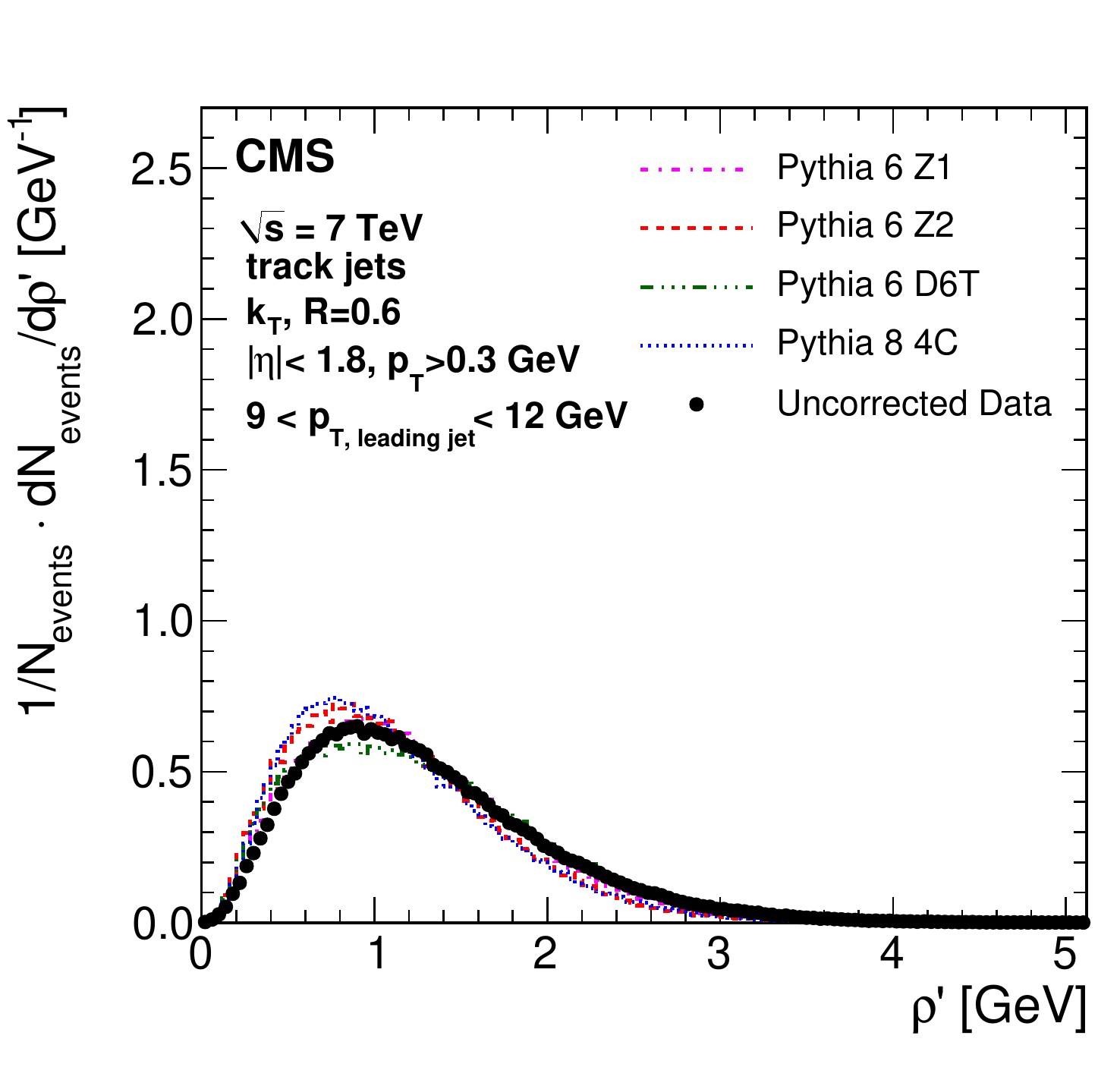}
  \includegraphics[width=\mygraphicswidth]{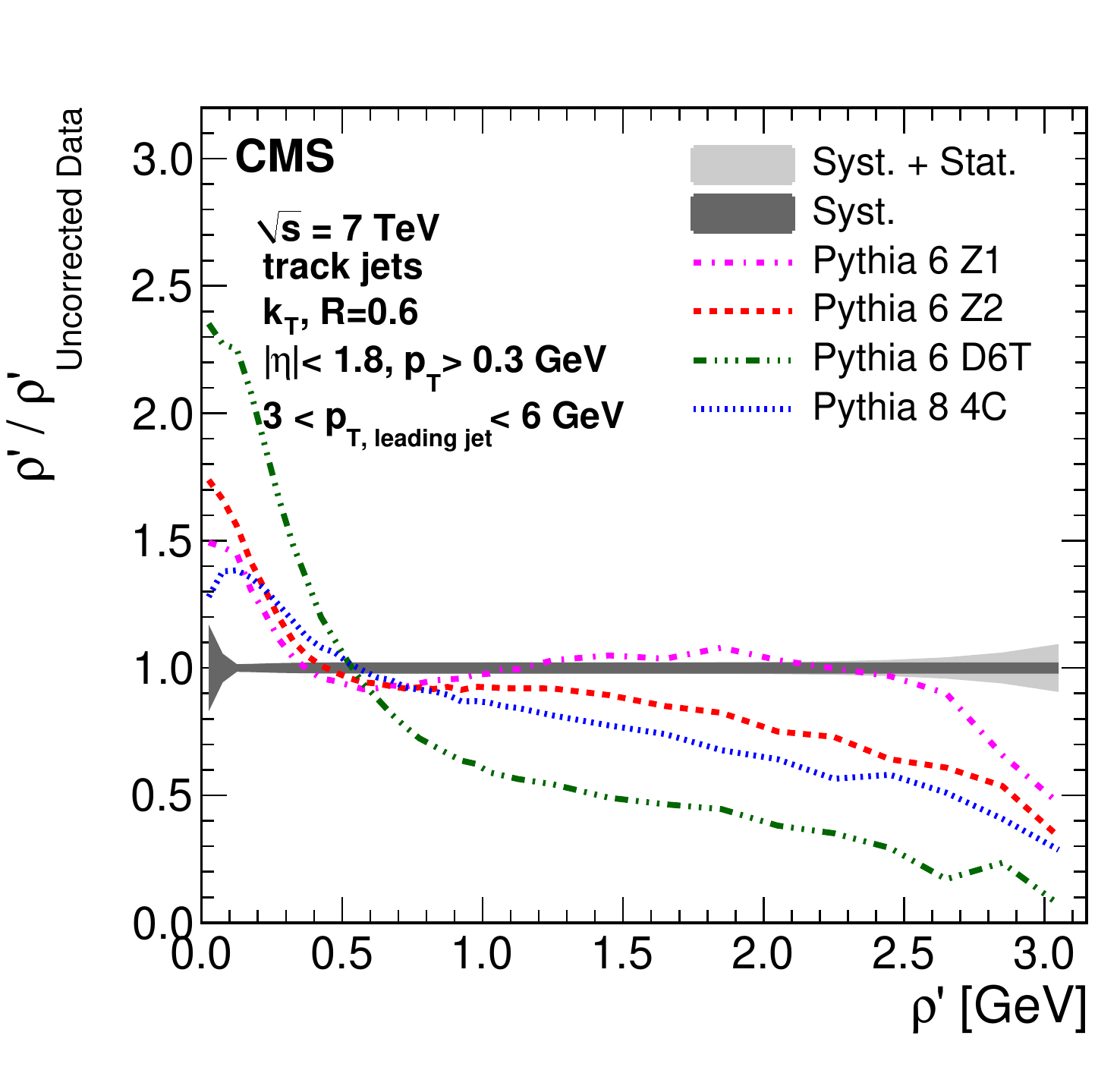}\hftwo%
  \includegraphics[width=\mygraphicswidth]{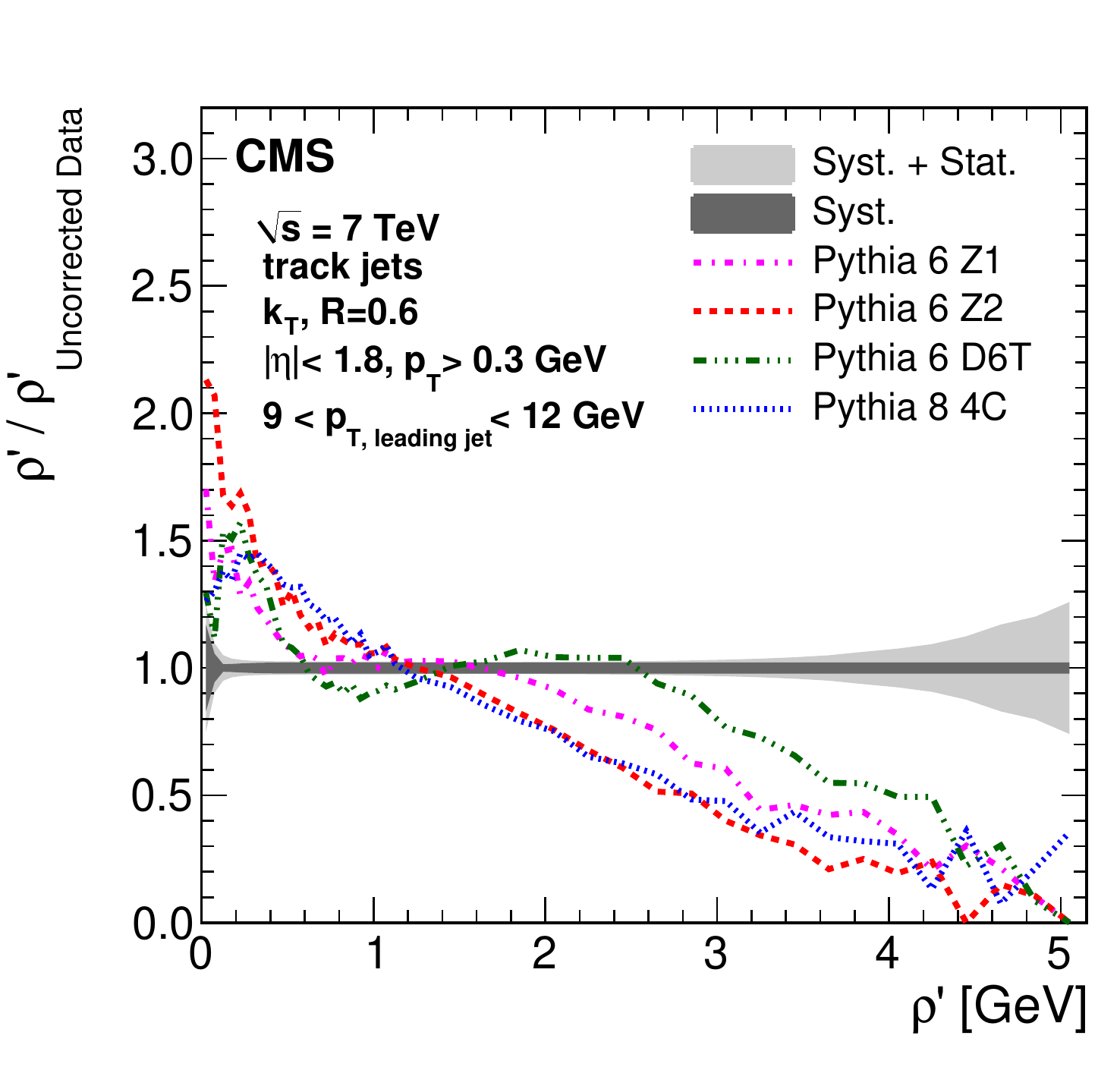}
  \caption{Uncorrected \rhop distributions in the two slices of
    leading track jet transverse momentum, $3 < p_\mathrm{T,leading} <
    6\GeV$ (left) and $9 < p_\mathrm{T,leading} < 12\GeV$ (right) at
    $\sqrt{s} = 7\TeV$. The reach in \rhop is different for the two
    slices in leading track jet \pt. The lower plots show the ratios
    of the different generator tunes to the reconstructed data. The
    dark grey shaded band corresponds to the systematic uncertainty
    and the light grey shaded band to the quadratic sum of the
    systematic and statistical uncertainty.}
  \label{Fig:kt6TrackJetsWithArearho_slices}
\end{figure}

Figure~\ref{Fig:kt6TrackJetsWithArearho_slices} shows as examples the
uncorrected \rhop distributions in the two slices of the leading jet
\pt of $3 < p_{\mathrm{T}\text{,leading}} < 6\GeV$ (left) and $9 <
p_{\mathrm{T}\text{,leading}} < 12\GeV$ (right) at $\sqrt{s} = 7\TeV$. Of
course, the additional binning in the hardness of the events
effectively limits the accessible range in \rhop as well. The observed
deviations of \PYTHIA~6 Z2 or \PYTHIA~8 4C from the measurements
remain similar when going from an inclusive view to slices in event
scale. In contrast to this, the comparison of \PYTHIA~6 D6T and Z1
relative to the data does change. This can be seen even more clearly
when concentrating on gross features of the distributions such as the
peak values, means, or widths, which depend visibly on the event
scale.

\begin{figure}
  \includegraphics[width=\mygraphicswidth]{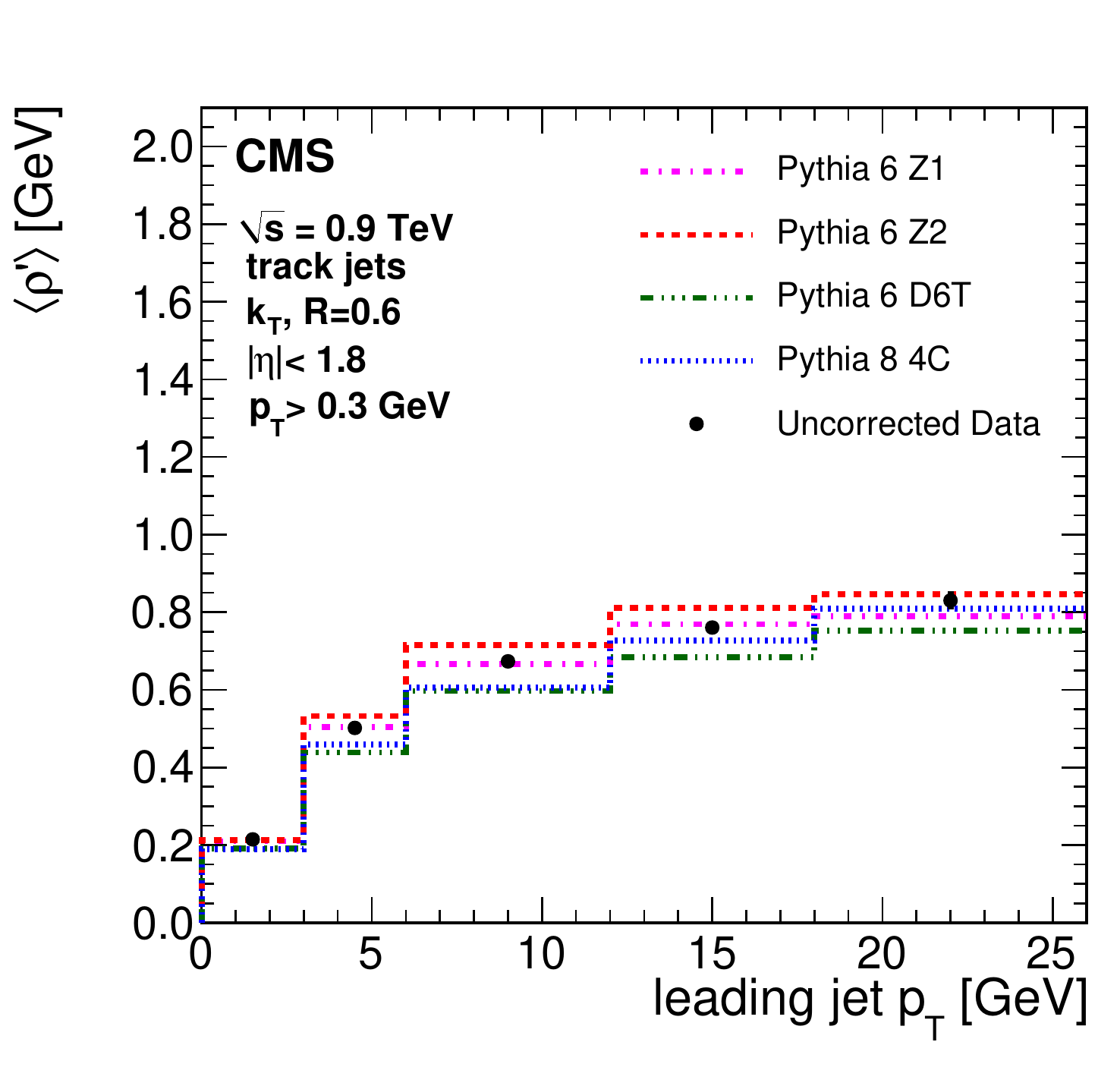}\hftwo%
  \includegraphics[width=\mygraphicswidth]{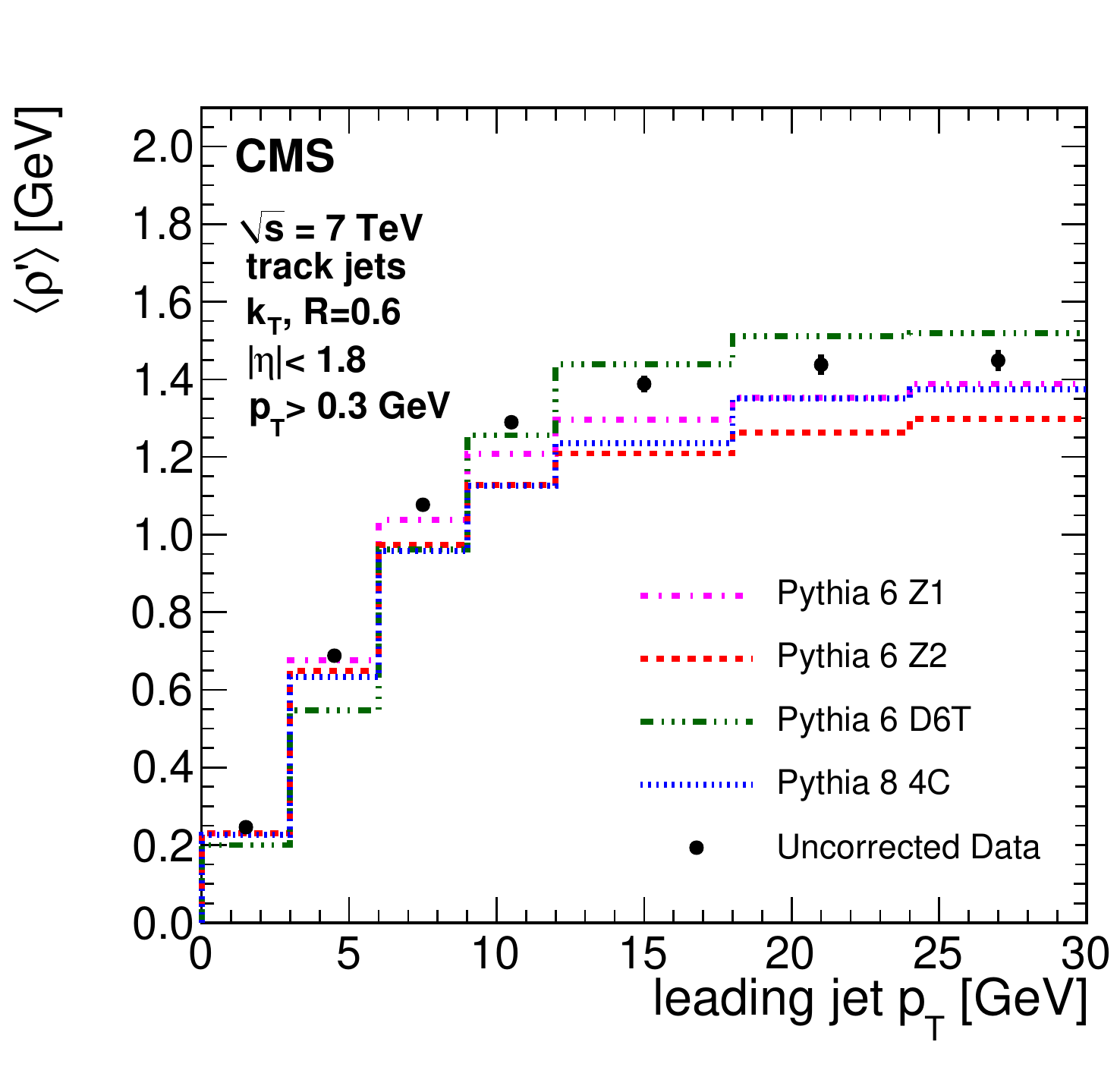}
  \caption{Mean values of the uncorrected \rhop distributions versus
    leading track jet transverse momentum at $\sqrt{s} = 0.9\TeV$
    (left) and $\sqrt{s} = 7\TeV$ (right) in comparison to the
    predictions by the different generator tunes. The error bars,
    which are mostly smaller than the symbol sizes, correspond to the
    quadratic sum of the systematic and statistical uncertainty.}
  \label{Fig:kt6TrackJetsWithArearho_means}
\end{figure}

For completeness Fig.~\ref{Fig:kt6TrackJetsWithArearho_means} presents
the mean values \rhopm for all slices possible at 0.9 and 7\TeV
centre-of-mass energy versus the leading jet \pt as event scale. In
accordance with expectations from similar UE studies in the
conventional approach, a steep rise at small event scales as well as a
saturation or plateau region at high scales are exhibited. The
increase of the UE activity with higher centre-of-mass energies is
visible from the heights of the plateau regions, which roughly
correspond to a ratio of $1.8$, in agreement with observations of a
ratio around $2$ for conventional observables
in~\cite{CMS-PAPERS-QCD-10-010}.

With respect to the tune comparisons at 0.9 and 7\TeV, \PYTHIA~8 4C
always undershoots the average UE activity as characterised by \rhopm,
\PYTHIA~6 Z1 changes from agreement with data to an underestimation,
Z2 from an overestimation to an underestimation, and D6T from a
systematic underestimation to an overestimation for event scales
larger than 10\GeV at 7\TeV centre-of-mass energy.

\subsection{Unfolded results}
\label{sec:unfoldedresults}

\begin{figure}
  \includegraphics[width=\mygraphicswidth]{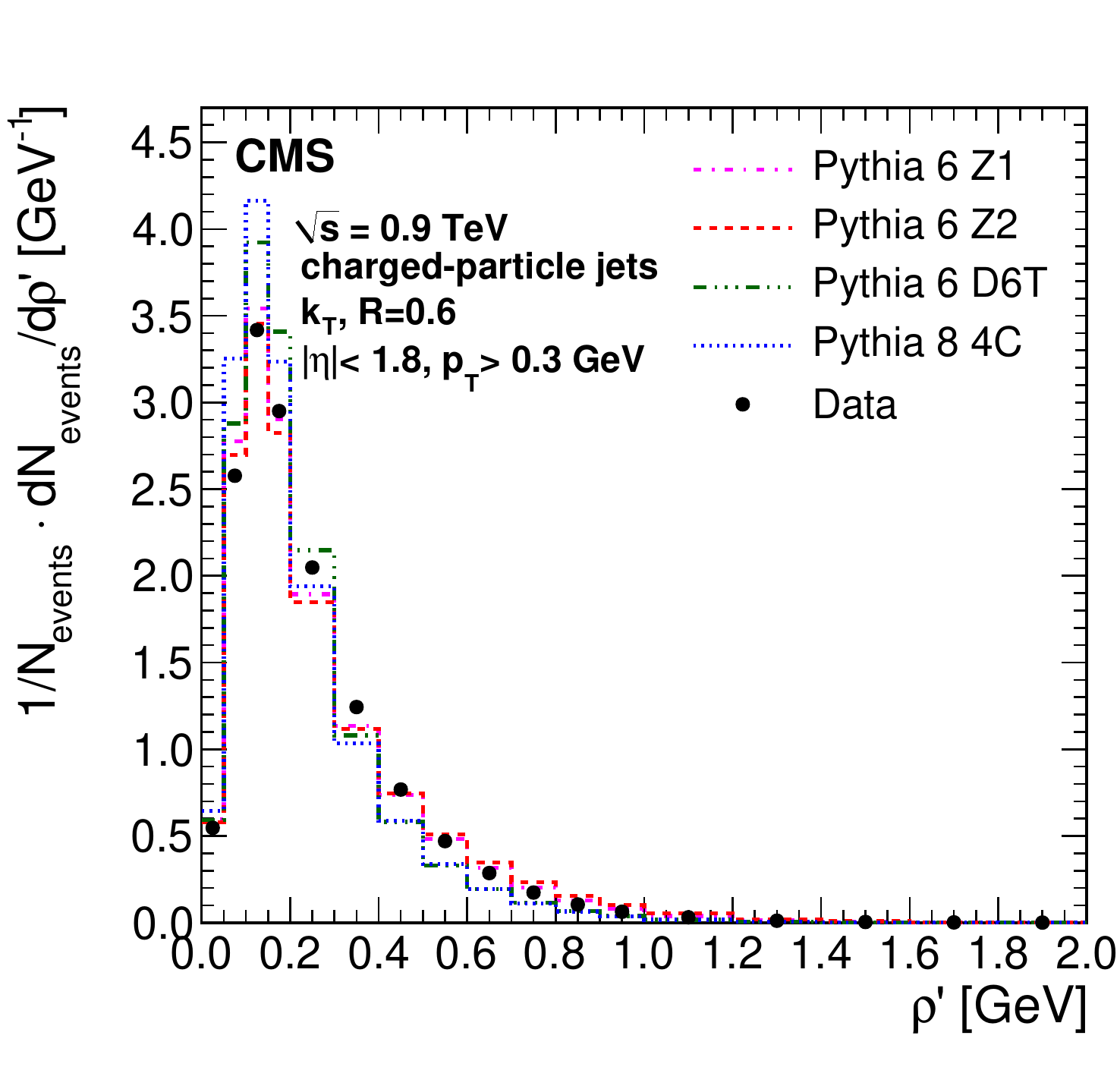}\hftwo%
  \includegraphics[width=\mygraphicswidth]{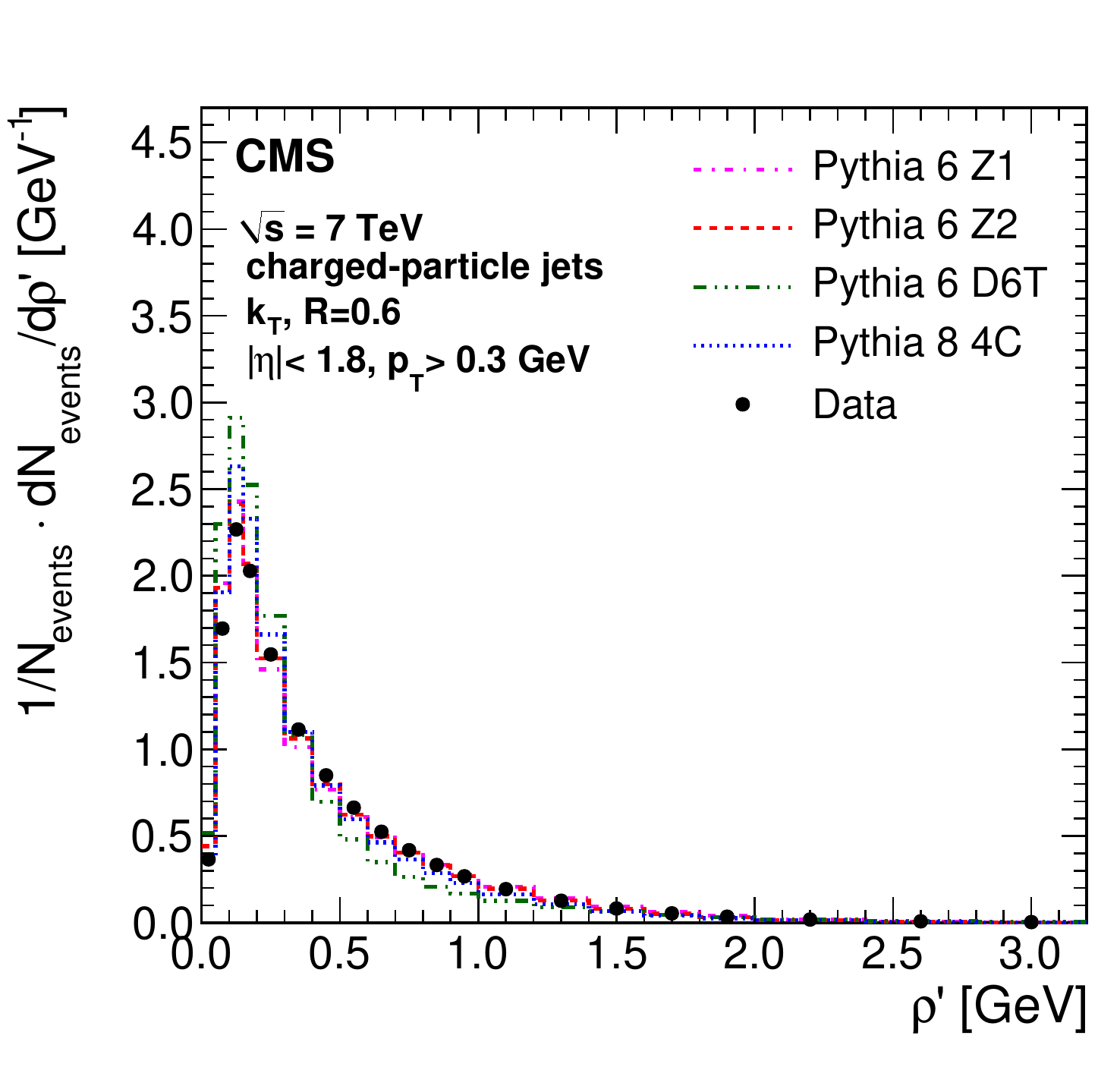}
  \includegraphics[width=\mygraphicswidth]{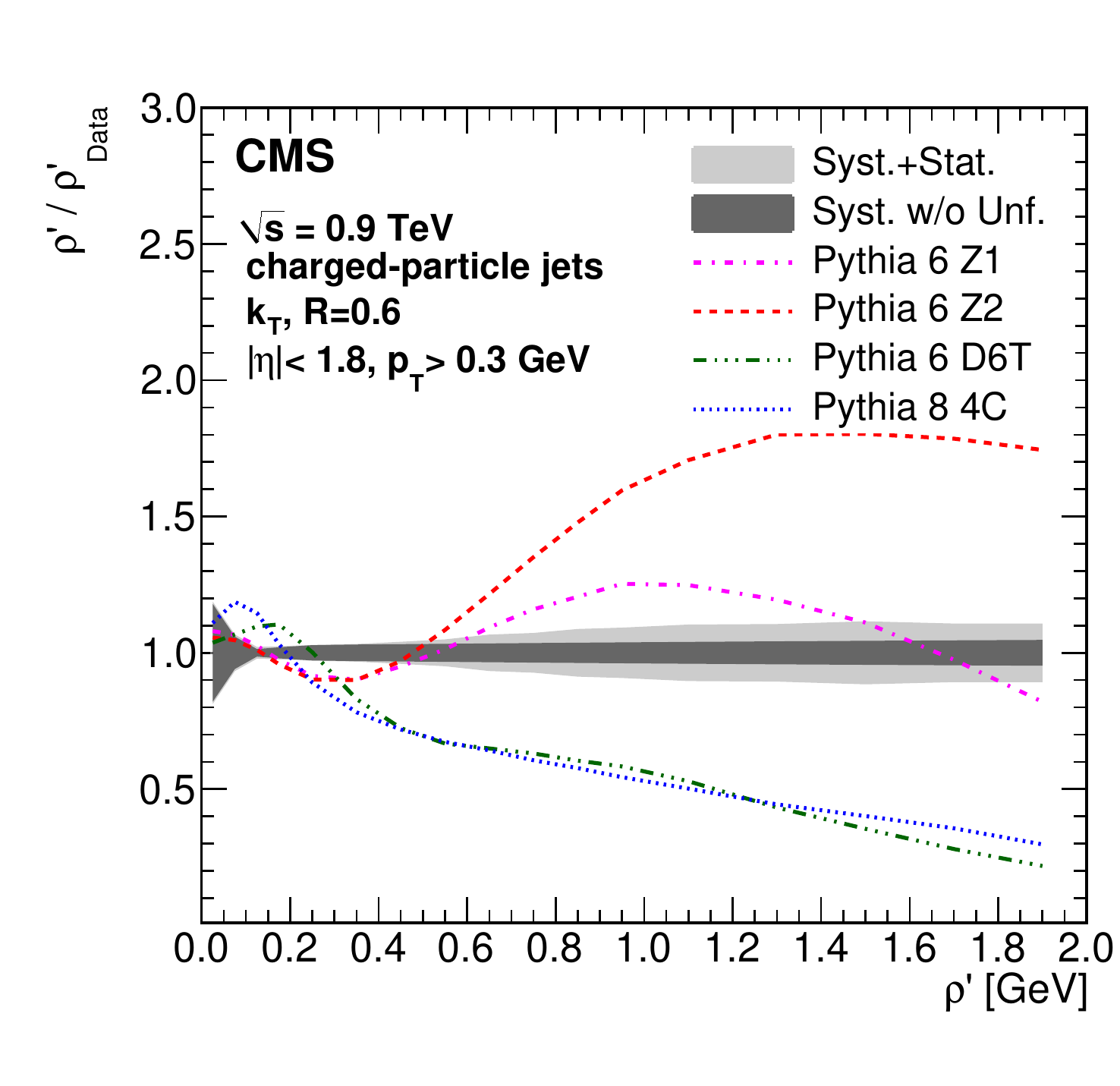}\hftwo%
  \includegraphics[width=\mygraphicswidth]{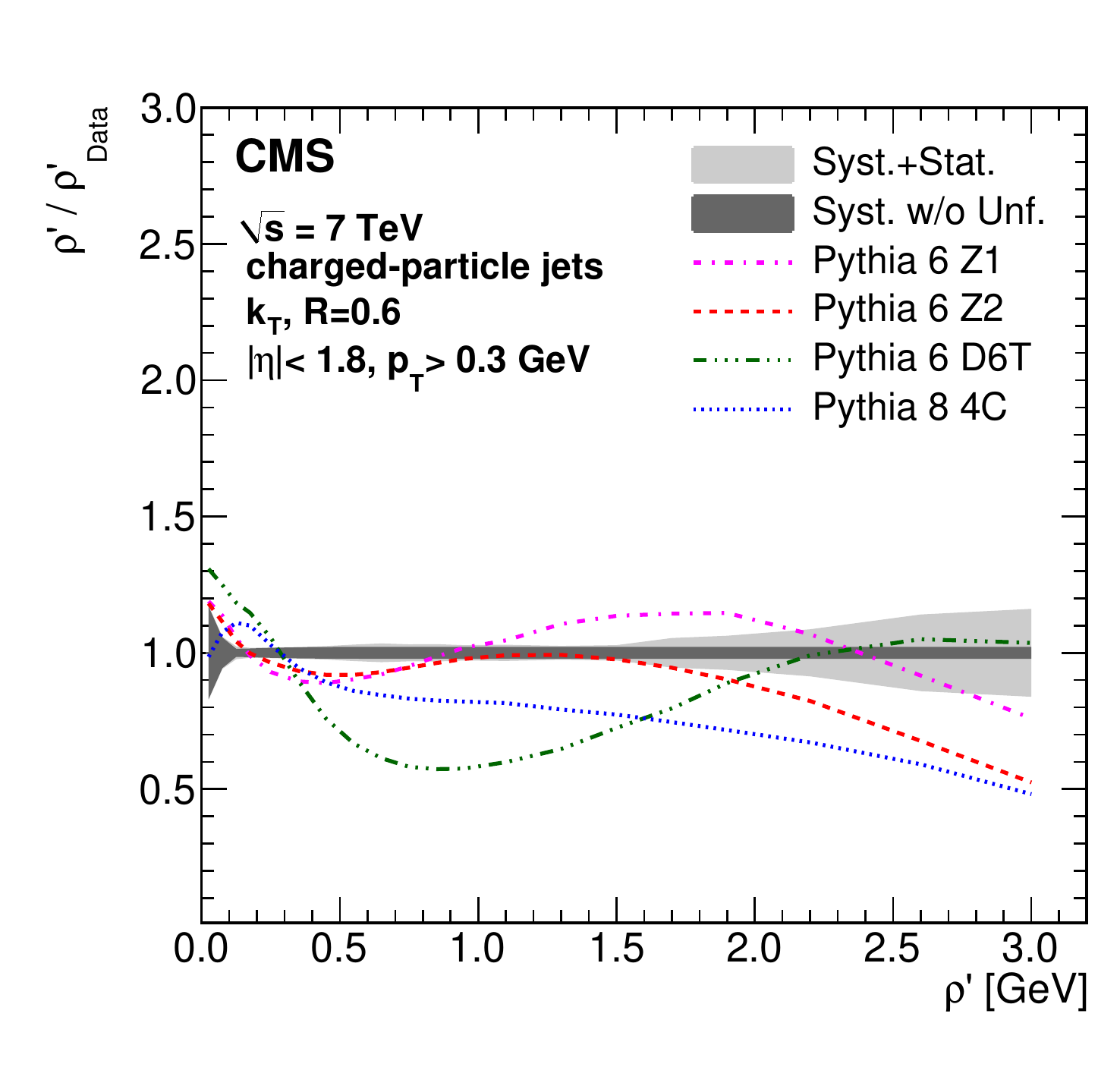}
  \caption{Unfolded inclusive \rhop distributions for data and
    simulation (upper row), and ratios of the \PYTHIA~6 tunes Z1, Z2,
    D6T, and the \PYTHIA~8 tune 4C relative to data (lower row) at
    $\sqrt{s} = 0.9\TeV$ (left) and $\sqrt{s} = 7\TeV$ (right). The
    quadratic difference between the total uncertainty, as given by
    the light grey band, and the dark grey band corresponds to the
    unfolding uncertainty, which inherently also comprises
    the statistical uncertainty.}
  \label{Fig:kt6TrackJetsWithArea_rho_inclusive_ufd}
\end{figure}

Figure~\ref{Fig:kt6TrackJetsWithArea_rho_inclusive_ufd} compares the
inclusive \rhop distributions, unfolded with the Bayesian method,
to the \PYTHIA~6 tunes Z1, Z2, D6T, and the \PYTHIA~8 tune 4C,
but this time at the level of stable charged particles. Because of the
response differences among the tunes, a substantial systematic
uncertainty is introduced by the unfolding, which is indicated in
Fig.~\ref{Fig:kt6TrackJetsWithArea_rho_inclusive_ufd} by the
difference between the dark and light grey shaded bands. Also the
range in \rhop had to be limited to $\rhop < 2.0\GeV$ for $\sqrt{s} =
0.9\TeV$ and $\rhop < 3.2\GeV$ for $\sqrt{s} = 7\TeV$ to ensure
the stability of the procedure.  Nevertheless, the
shape of the \rhop distributions is rather well preserved during the
unfolding process and the same conclusions can be drawn as from the
comparison of the uncorrected observable.

\begin{figure}
  \includegraphics[width=\mygraphicswidth]{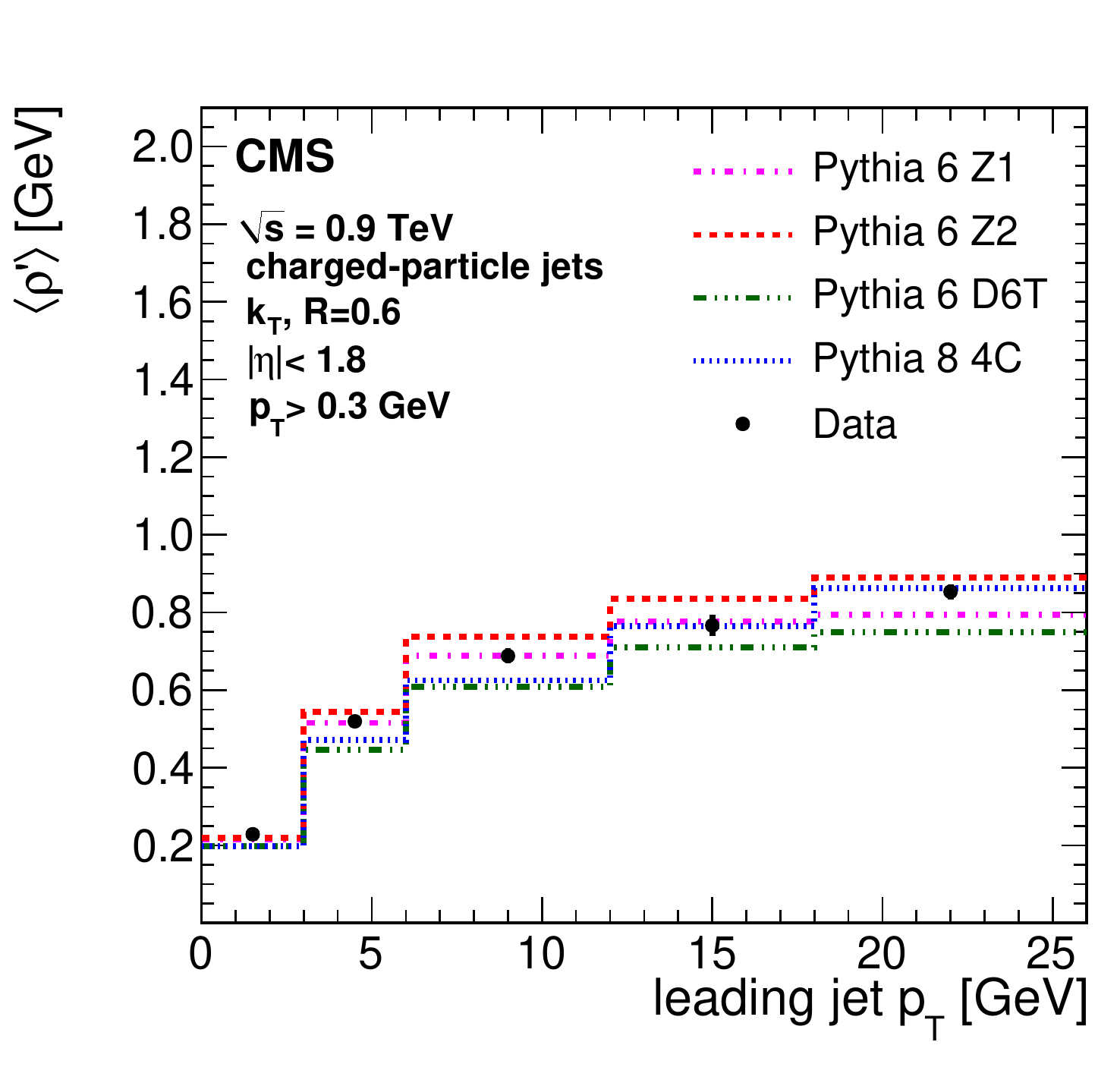}\hftwo%
  \includegraphics[width=\mygraphicswidth]{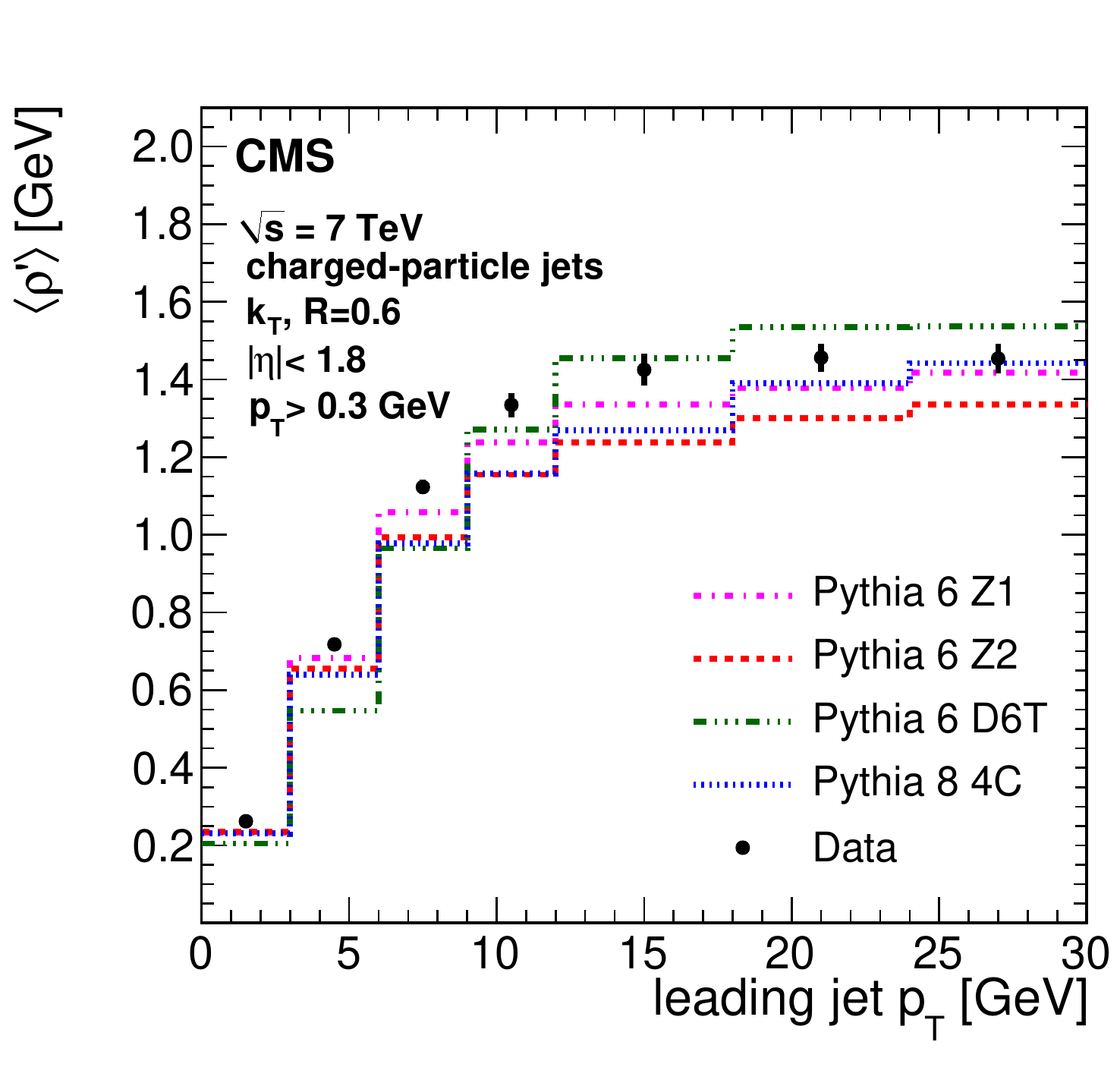}
  \caption{Mean values of the corrected \rhop distributions versus
    leading charged-particle jet transverse momentum at
    $\sqrt{s} = 0.9\TeV$ (left) and $\sqrt{s} = 7\TeV$ (right) in
    comparison to the predictions by the different generator tunes.
    The \rhop distributions in each slice are unfolded with the
    Bayesian method. The error bars, which are mostly
    smaller than the symbol sizes, correspond to the total
    uncertainty.
  }
  \label{Fig:kt6TrackJetsWithArearho_means_ufd}
\end{figure}

For the purpose of deriving the event scale dependence of \rhopm, the
\rhop distribution in each slice of jet \pt must be unfolded
independently using separate response matrices. The result is
presented in Fig.~\ref{Fig:kt6TrackJetsWithArearho_means_ufd} where
the error bars are dominated by the uncertainty introduced through the
unfolding procedure. Again, the observations are consistent with the
uncorrected case as shown in
Fig.~\ref{Fig:kt6TrackJetsWithArearho_means} and the ratio of the
plateau heights roughly corresponds to a factor of $1.8$ between 0.9 and
7\TeV centre-of-mass energy.

\section{Summary}
\label{sec:summary}

The jet-area/median approach to measuring the underlying event has
been studied for the first time in a collider experiment at the two
centre-of-mass energies of $0.9$ and $7\TeV$ with the CMS
detector. The measured distributions of the observable \rhop, based on
this approach, are unfolded for detector effects and compared to
predictions of several Monte Carlo event generator tunes before and
after detector simulation. The substantial discrepancies observed
among the various predictions and also between the predictions and the
data demonstrate the sensitivity of the method and indicate the need
for improved tunes at both centre-of-mass energies. None of the
examined models describe the data satisfactorily.

Overall, \PYTHIA~6 Z1 gives the best description of the data with some
residual underestimation at $\sqrt{s} = 7\TeV$. \PYTHIA~6 Z2 varies
from severely overshooting the data at $0.9\TeV$ to falling short of
the data at $7\TeV$, hinting at an inadequate setting of the
$\sqrt{s}$ dependence for the UE model. \PYTHIA~8 4C almost always
underestimates the UE activity, while \PYTHIA~6 D6T does so only at
0.9\TeV or at small event scales but then rises too steeply with
increasing hardness of the events. The general pattern of deviations
from data by the considered \PYTHIA tunes is similar to that observed
with the conventional approach~\cite{CMS-PAPERS-QCD-10-010}.

The mean \rhopm has also been investigated as a function of the
transverse momentum of the leading jet. In agreement with the
conventional analysis, a steep rise of the UE activity with increasing
leading jet transverse momentum up to about $10\GeV$ is observed. For
higher transverse momenta a plateau is reached. The ratio of the UE
activity in this saturation region at $7\TeV$ to that at $0.9\TeV$ is
approximately $1.8$, which is close to the ratios of around $2$
measured with the conventional observables.

In conclusion, the new observable \rhop based on the jet-area/median
approach has been demonstrated to be sensitive to soft hadronic
activity and offers an alternative view of the UE\@. The method is not
restricted to minimum-bias events as examined here but can also be
applied to different event topologies.

\section*{Acknowledgments}
\label{sec:ack}

We thank G.~Salam and S.~Sapeta for help in defining the modified
observable \rhop.

\hyphenation{Bundes-ministerium Forschungs-gemeinschaft
  Forschungs-zentren} We congratulate our colleagues in the CERN
accelerator departments for the excellent performance of the LHC
machine. We thank the technical and administrative staff at CERN and
other CMS institutes. This work was supported by the Austrian Federal
Ministry of Science and Research; the Belgium Fonds de la Recherche
Scientifique, and Fonds voor Wetenschappelijk Onderzoek; the Brazilian
Funding Agencies (CNPq, CAPES, FAPERJ, and FAPESP); the Bulgarian
Ministry of Education and Science; CERN; the Chinese Academy of
Sciences, Ministry of Science and Technology, and National Natural
Science Foundation of China; the Colombian Funding Agency
(COLCIENCIAS); the Croatian Ministry of Science, Education and Sport;
the Research Promotion Foundation, Cyprus; the Ministry of Education
and Research, Recurrent financing contract SF0690030s09 and European
Regional Development Fund, Estonia; the Academy of Finland, Finnish
Ministry of Education and Culture, and Helsinki Institute of Physics;
the Institut National de Physique Nucl\'eaire et de Physique des
Particules~/~CNRS, and Commissariat \`a l'\'Energie Atomique et aux
\'Energies Alternatives~/~CEA, France; the Bundesministerium f\"ur
Bildung und Forschung, Deutsche Forschungsgemeinschaft, and
Helmholtz-Gemeinschaft Deutscher Forschungszentren, Germany; the
General Secretariat for Research and Technology, Greece; the National
Scientific Research Foundation, and National Office for Research and
Technology, Hungary; the Department of Atomic Energy and the
Department of Science and Technology, India; the Institute for Studies
in Theoretical Physics and Mathematics, Iran; the Science Foundation,
Ireland; the Istituto Nazionale di Fisica Nucleare, Italy; the Korean
Ministry of Education, Science and Technology and the World Class
University program of NRF, Korea; the Lithuanian Academy of Sciences;
the Mexican Funding Agencies (CINVESTAV, CONACYT, SEP, and UASLP-FAI);
the Ministry of Science and Innovation, New Zealand; the Pakistan
Atomic Energy Commission; the Ministry of Science and Higher Education
and the National Science Centre, Poland; the Funda\c{c}\~ao para a
Ci\^encia e a Tecnologia, Portugal; JINR (Armenia, Belarus, Georgia,
Ukraine, Uzbekistan); the Ministry of Education and Science of the
Russian Federation, the Federal Agency of Atomic Energy of the Russian
Federation, Russian Academy of Sciences, and the Russian Foundation
for Basic Research; the Ministry of Science and Technological
Development of Serbia; the Secretar\'{\i}a de Estado de
Investigaci\'on, Desarrollo e Innovaci\'on and Programa
Consolider-Ingenio 2010, Spain; the Swiss Funding Agencies (ETH Board,
ETH Zurich, PSI, SNF, UniZH, Canton Zurich, and SER); the National
Science Council, Taipei; the Scientific and Technical Research Council
of Turkey, and Turkish Atomic Energy Authority; the Science and
Technology Facilities Council, UK; the US Department of Energy, and
the US National Science Foundation.

Individuals have received support from the Marie-Curie programme and
the European Research Council (European Union); the Leventis
Foundation; the A. P. Sloan Foundation; the Alexander von Humboldt
Foundation; the Belgian Federal Science Policy Office; the Fonds pour
la Formation \`a la Recherche dans l'Industrie et dans l'Agriculture
(FRIA-Belgium); the Agentschap voor Innovatie door Wetenschap en
Technologie (IWT-Belgium); the Council of Science and Industrial
Research, India; the Compagnia di San Paolo (Torino); and the HOMING
PLUS programme of Foundation for Polish Science, cofinanced from
European Union, Regional Development Fund.

\bibliography{auto_generated}

\cleardoublepage \appendix\section{The CMS Collaboration \label{app:collab}}\begin{sloppypar}\hyphenpenalty=5000\widowpenalty=500\clubpenalty=5000\textbf{Yerevan Physics Institute,  Yerevan,  Armenia}\\*[0pt]
S.~Chatrchyan, V.~Khachatryan, A.M.~Sirunyan, A.~Tumasyan
\vskip\cmsinstskip
\textbf{Institut f\"{u}r Hochenergiephysik der OeAW,  Wien,  Austria}\\*[0pt]
W.~Adam, T.~Bergauer, M.~Dragicevic, J.~Er\"{o}, C.~Fabjan, M.~Friedl, R.~Fr\"{u}hwirth, V.M.~Ghete, J.~Hammer\cmsAuthorMark{1}, N.~H\"{o}rmann, J.~Hrubec, M.~Jeitler, W.~Kiesenhofer, V.~Kn\"{u}nz, M.~Krammer, D.~Liko, I.~Mikulec, M.~Pernicka$^{\textrm{\dag}}$, B.~Rahbaran, C.~Rohringer, H.~Rohringer, R.~Sch\"{o}fbeck, J.~Strauss, A.~Taurok, F.~Teischinger, P.~Wagner, W.~Waltenberger, G.~Walzel, E.~Widl, C.-E.~Wulz
% ...
\vskip\cmsinstskip
\dag:~Deceased\\
1:~~Also at CERN, European Organization for Nuclear Research, Geneva, Switzerland\\
% ...

\end{sloppypar}
\end{document}